\documentclass[twocolumn]{aastex701}
\usepackage{amsmath} 
\usepackage{hyperref}

\def\mailto#1{\href{mailto:#1}{#1}}

\begin{document}

\title{No Detectable One-halo Galactic Conformity Signal with Halo-mass Estimates Consistent with Weak-lensing Constraints\footnote{This is the seventeenth paper in the ``From Halos to Galaxies'' series.}
}

\correspondingauthor{Yingjie Peng} 
\email{yjpeng@pku.edu.cn}

\author[gname=Dingyi,sname=Zhao,0009-0001-1564-3944]{Dingyi Zhao}
\email{zhaodingyi@pku.edu.cn} 
\affiliation{Kavli Institute for Astronomy and Astrophysics, Peking University, 5 Yiheyuan Road, Beijing 100871, People's Republic of China; \mailto{yjpeng@pku.edu.cn}}
\affiliation{Department of Astronomy, School of Physics, Peking University, 5 Yiheyuan Road, Beijing 100871, People's Republic of China}

\author[gname=Yingjie,sname=Peng,0000-0003-0939-9671]{Yingjie Peng}
\email{yjpeng@pku.edu.cn} 
\affiliation{Kavli Institute for Astronomy and Astrophysics, Peking University, 5 Yiheyuan Road, Beijing 100871, People's Republic of China; \mailto{yjpeng@pku.edu.cn}}
\affiliation{Department of Astronomy, School of Physics, Peking University, 5 Yiheyuan Road, Beijing 100871, People's Republic of China}

\author[gname=Luis C.,sname=Ho,0000-0001-6947-5846]{Luis C. Ho}
\email{lho.pku@gmail.com} 
\affiliation{Kavli Institute for Astronomy and Astrophysics, Peking University, 5 Yiheyuan Road, Beijing 100871, People's Republic of China; \mailto{yjpeng@pku.edu.cn}}
\affiliation{Department of Astronomy, School of Physics, Peking University, 5 Yiheyuan Road, Beijing 100871, People's Republic of China}

\author[gname=Houjun,sname=Mo,0000-0001-5356-2419]{Houjun Mo}
\email{hjmo@umass.edu} 
\affiliation{Department of Astronomy, University of Massachusetts, Amherst, MA 01003, USA}

\author[gname=Cheqiu,sname=Lyu,0009-0000-7307-6362]{Cheqiu Lyu}
\email{lyucq@ustc.edu.cn} 
\affiliation{Department of Astronomy, University of Science and Technology of China, Hefei 230026, People's Republic of China}
\affiliation{School of Astronomy and Space Science, University of Science and Technology of China, Hefei 230026, People's Republic of China}

\author[gname=Kai,sname=Wang,0000-0002-3775-0484]{Kai Wang}
\email{wkcosmology@gmail.com} 
\affiliation{Institute for Computational Cosmology, Department of Physics, Durham University, South Road, Durham DH1 3LE, UK}
\affiliation{Centre for Extragalactic Astronomy, Department of Physics, Durham University, South Road, Durham DH1 3LE, UK}

\author[gname=Yijun,sname=Wang,]{Yijun Wang}
\email{wangyijun@nju.edu.cn} 
\affiliation{School of Astronomy and Space Science, Nanjing University, Nanjing 210093, People's Republic of China}
\affiliation{Key Laboratory of Modern Astronomy and Astrophysics, Nanjing University, Ministry of Education, Nanjing 210093, People's Republic of China}

\author[gname=Tao,sname=Wang,]{Tao Wang}
\email{taowang@nju.edu.cn} 
\affiliation{School of Astronomy and Space Science, Nanjing University, Nanjing 210093, People's Republic of China}
\affiliation{Key Laboratory of Modern Astronomy and Astrophysics, Nanjing University, Ministry of Education, Nanjing 210093, People's Republic of China}

\author[gname=Jing,sname=Dou,0000-0002-6961-6378]{Jing Dou}
\email{doujing@nao.cas.cn} 
\affiliation{National Astronomical Observatories, Chinese Academy of Sciences, Beijing 100101, People's Republic of China}

\author[gname=Zeyu,sname=Gao,0000-0002-0182-1973]{Zeyu Gao}
\email{zygao@stu.pku.edu.cn} 
\affiliation{Department of Astronomy, School of Physics, Peking University, 5 Yiheyuan Road, Beijing 100871, People's Republic of China}
\affiliation{Kavli Institute for Astronomy and Astrophysics, Peking University, 5 Yiheyuan Road, Beijing 100871, People's Republic of China; \mailto{yjpeng@pku.edu.cn}}

\author[gname=Qiusheng,sname=Gu,0000-0002-3890-3729]{Qiusheng Gu}
\email{qsgu@nju.edu.cn} 
\affiliation{School of Astronomy and Space Science, Nanjing University, Nanjing 210093, People's Republic of China}

\author[gname=Yukun,sname=Liu,]{Yukun Liu}
\email{2501110276@stu.pku.edu.cn} 
\affiliation{Department of Astronomy, School of Physics, Peking University, 5 Yiheyuan Road, Beijing 100871, People's Republic of China}
\affiliation{Kavli Institute for Astronomy and Astrophysics, Peking University, 5 Yiheyuan Road, Beijing 100871, People's Republic of China; \mailto{yjpeng@pku.edu.cn}}

\author[gname=Roberto,sname=Maiolino,0000-0002-4985-3819]{Roberto Maiolino}
\email{r.maiolino@mrao.cam.ac.uk} 
\affiliation{Cavendish Laboratory, University of Cambridge, 19 J.J. Thomson Avenue, Cambridge CB3 0HE, UK}
\affiliation{Kavli Institute for Cosmology, University of Cambridge, Madingley Road, Cambridge CB3 0HA, UK}
\affiliation{Department of Physics and Astronomy, University College London, Gower Street, London WC1E 6BT, UK}

\author[gname=Alvio,sname=Renzini,0000-0002-7093-7355]{Alvio Renzini}
\email{alvio.renzini@oapd.inaf.it} 
\affiliation{INAF--Osservatorio Astronomico di Padova, Vicolo dell'Osservatorio 5, I-35122 Padova, Italy}

\author[gname=Bitao,sname=Wang,0000-0002-6137-6007]{Bitao Wang}
\email{bt-wang@pku.edu.cn} 
\affiliation{School of Physics and Electronics, Hunan University, Changsha, Hunan 410082, People's Republic of China}

\author[gname=Yu-Chen,sname=Wang,0000-0002-8429-7088]{Yu-Chen Wang}
\email{ycwang-astro@stu.pku.edu.cn} 
\affiliation{Department of Astronomy, School of Physics, Peking University, 5 Yiheyuan Road, Beijing 100871, People's Republic of China}
\affiliation{Kavli Institute for Astronomy and Astrophysics, Peking University, 5 Yiheyuan Road, Beijing 100871, People's Republic of China; \mailto{yjpeng@pku.edu.cn}}

\author[gname=Bingxiao,sname=Xu,]{Bingxiao Xu}
\email{bxu6@pku.edu.cn} 
\affiliation{Kavli Institute for Astronomy and Astrophysics, Peking University, 5 Yiheyuan Road, Beijing 100871, People's Republic of China; \mailto{yjpeng@pku.edu.cn}}

\author[gname=Feng,sname=Yuan,0000-0003-3564-6437]{Feng Yuan}
\email{fyuan@shao.ac.cn} 
\affiliation{Center for Astronomy and Astrophysics and Department of Physics, Fudan University, Shanghai 200438, People's Republic of China}

\author[gname=Kunyao,sname=Zhao,]{Kunyao Zhao}
\email{zky_1005@stu.pku.edu.cn} 
\affiliation{Department of Astronomy, School of Physics, Peking University, 5 Yiheyuan Road, Beijing 100871, People's Republic of China}
\affiliation{Kavli Institute for Astronomy and Astrophysics, Peking University, 5 Yiheyuan Road, Beijing 100871, People's Republic of China; \mailto{yjpeng@pku.edu.cn}}

\author[gname=Xingye,sname=Zhu,0000-0002-9529-1044]{Xingye Zhu}
\email{zhuxy.astro@gmail.com} 
\affiliation{Department of Astronomy, School of Physics, Peking University, 5 Yiheyuan Road, Beijing 100871, People's Republic of China}
\affiliation{Kavli Institute for Astronomy and Astrophysics, Peking University, 5 Yiheyuan Road, Beijing 100871, People's Republic of China; \mailto{yjpeng@pku.edu.cn}}

\begin{abstract}
One-halo galactic conformity is the tendency for satellites in halos with quenched centrals to have lower star-formation activity than those in halos with star-forming centrals at fixed halo mass. It is an important probe of the galaxy--halo connection and halo-wide quenching processes that may couple central and satellite evolution. However, its existence remains controversial, because conformity must be measured at fixed halo mass, while halo masses are difficult to estimate accurately. 
\added{In this Letter, we measure one-halo conformity in SDSS using five stellar-mass-complete samples and three halo-mass estimates: an ML estimate whose star-forming and quenched stellar mass--halo mass relations (SHMRs) agree with independent weak-lensing constraints, and two conventional abundance-matching (AM) estimates.}
We quantify conformity as the difference in median $\log({\rm sSFR})$ between satellites of star-forming and quenched centrals, using both satellite-level and halo-level statistics. The two AM estimates produce strong positive conformity signals, consistent with previous AM-based measurements, but these signals are not reproduced with the ML halo masses. For the halo-level statistic, the representative AM-based signals are $+0.38\pm0.04$ dex and $+0.23\pm0.04$ dex for the luminosity-ranking and mass-ranking AM halo masses, detected relative to no conformity at about $10\sigma$ and $6\sigma$, respectively. In contrast, the ML result is consistent with no conformity, $+0.00\pm0.03$ dex; the satellite-level statistic gives a similar result. 
\added{Thus, with halo-mass estimates consistent with weak-lensing constraints, we find no detectable one-halo conformity signal in the present SDSS sample, suggesting that the strong AM-based signal is largely driven by halo-mass estimation biases.}
\end{abstract}



\section{Introduction} \label{sec:intro}
The quenching of star formation in galaxies is a central problem in galaxy evolution. It is well established that a galaxy's star formation rate (SFR) correlates strongly with both its stellar mass and its environment (e.g., \citealt{2004ApJ...615L.101B,2006MNRAS.373..469B,2007MNRAS.376..841V,2008MNRAS.387...79V,2010ApJ...721..193P,2012ApJ...757....4P,2013MNRAS.428.3306W,2014MNRAS.441..599B,2020MNRAS.491L..51P,2023ApJ...959....5L}). A more subtle environmental phenomenon is one-halo galactic conformity, namely the tendency, at fixed halo mass, for satellites in halos with quenched central galaxies to have lower star-formation activity than satellites in halos with star-forming central galaxies. Such a signal would provide an important probe of the galaxy--halo connection and of halo-wide quenching processes, such as halo quenching and active galactic nucleus (AGN) feedback, that may couple the evolution of central and satellite galaxies.

One-halo conformity was first reported by \citet{2006MNRAS.366....2W}, who found that quenched central galaxies in SDSS groups tend to host a higher fraction of quenched satellites than star-forming centrals, even at apparently fixed halo mass. Subsequent observational studies, using either group catalogs or isolation-based methods, also reported similar signals in star formation and morphology (e.g., \citealt{2008MNRAS.389...86A,2015ApJ...800...24K,2020MNRAS.492.2722O}).

However, whether this signal is genuinely present has remained controversial. Since one-halo conformity is defined through comparisons at fixed halo mass, the measurement depends critically on the accuracy of halo-mass estimates. \citet{2012MNRAS.424.2574W} pointed out that quenched and star-forming centrals can have different halo-mass distributions, so inaccurate halo-mass estimates can mix halos of different true masses and produce an apparent conformity signal. More generally, mock-catalog and simulation-based studies have shown that the inferred one-halo conformity signal can be strongly affected by systematics in group finding and, especially, in halo-mass assignment \citep{2015MNRAS.452..444C,2018MNRAS.480.2031C}. These results emphasize that reliable halo-mass estimates are essential for one-halo conformity measurements.

In this Letter, we address this key halo-mass uncertainty using the machine-learning (ML) halo-mass estimate developed by \citet{2025ApJ...979...42Z}. Compared with standard abundance-matching (AM) halo masses, this method substantially reduces halo-mass scatter and yields stellar-to-halo mass relations for both star-forming and quenched galaxies that are consistent with weak-lensing measurements \citep{2016MNRAS.457.3200M,2021A&A...653A..82B}. Using a set of stellar-mass-complete SDSS samples, we compare the one-halo conformity signal inferred from the ML halo masses with that obtained from the conventional mass-ranking and luminosity-ranking AM halo masses. We perform this comparison using both the commonly used satellite-level statistic and a halo-level statistic. We show that the strong conformity signal obtained with AM halo masses is not reproduced when the same analysis is repeated using the ML halo masses.

\section{Data and Method}
\label{sec:data}

\subsection{Galaxy Sample}
\label{subsec:data_sample}

We draw our galaxy sample from the Sloan Digital Sky Survey Data Release 7 (SDSS DR7; \citealt{2009ApJS..182..543A}). Group membership is taken from the updated group catalog of \citet{2007ApJ...671..153Y}, which was constructed from the NYU-VAGC DR7 \citep{2005AJ....129.2562B} using an adaptive halo-based group finder \citep{2005MNRAS.356.1293Y}. We adopt the catalog version based on \texttt{model} magnitudes. In each group, the most massive galaxy is defined as the central galaxy, and all other group members are treated as satellites.

Stellar masses, star formation rates (SFRs), and redshifts are taken from the MPA-JHU DR7 release of spectral measurements.\footnote{\url{https://wwwmpa.mpa-garching.mpg.de/SDSS/DR7/}} Stellar masses are obtained through photometric fitting using a Bayesian methodology \citep{2003MNRAS.341...33K}. SFRs are derived from H$\alpha$ emission for star-forming galaxies and, when emission lines are unreliable, are estimated using a D4000-based calibration \citep{2004MNRAS.351.1151B}; the resulting SFRs are then aperture-corrected using the photometry outside the SDSS fiber \citep{2007ApJS..173..267S}.

Following \citet{2015ApJ...801L..29R}, we adopt their star-forming main sequence and define star-forming/quenched galaxies using an offset of 0.7 dex below that relation. Equivalently, galaxies are classified as star-forming (SF) if
\begin{equation}
    \log \left( \frac{{\rm SFR}}{M_{\odot}\,{\rm yr}^{-1}} \right)
    >
    0.76\,\log \left( \frac{M_*}{M_{\odot}} \right) - 8.34,
\end{equation}
and as quenched (Q) otherwise.

\subsection{Halo Mass Estimation}
\label{subsec:halo_mass}

A central goal of this work is to re-measure one-halo conformity using halo-mass estimates that are less affected by the known biases of traditional abundance matching (AM). \added{In particular, the AM halo masses used here exhibit a population-dependent systematic offset relative to galaxy--galaxy weak-lensing measurements: at fixed stellar mass, they tend to overestimate the halo masses of star-forming centrals and underestimate those of quenched centrals \citep{2025ApJ...979...42Z}. Such a bias is especially important for conformity measurements, because it can mix halos of different true masses when star-forming- and quenched-central systems are compared at fixed inferred halo mass.}

We therefore adopt as our fiducial estimate the machine-learning (ML) halo mass developed by \citet{2025ApJ...979...42Z}. This method uses multiple properties of galaxy groups as input features and is trained on a mock catalog constructed from the L-GALAXIES semi-analytic model \citep{2015MNRAS.451.2663H}. After calibration to SDSS observations, the ML model is implemented with \textsc{XGBoost} \citep{2016arXiv160302754C} and applied to the Yang et al.\ group catalog.

As shown by \citet{2025ApJ...979...42Z}, in the test set, this ML method reduces the scatter in halo-mass estimation by approximately one-third relative to standard AM estimates, and substantially reduces the systematic offset between star-forming and quenched populations. For observational data, it yields a halo mass function consistent with simulations and stellar-to-halo mass relations for both star-forming and quenched galaxies that agree with independent weak-lensing measurements \citep{2016MNRAS.457.3200M,2021A&A...653A..82B}. \added{This agreement is particularly important for the present analysis, because the population-dependent halo-mass bias seen in the AM estimates can directly generate or enhance an apparent conformity signal.}

For comparison, we also use two standard AM halo-mass estimates, namely the mass-ranking AM halo mass and the luminosity-ranking AM halo mass, both taken directly from the \citet{2007ApJ...671..153Y} group catalog.

\begin{table}[t]
\centering
\caption{Definition of the five complete samples.}
\label{tab:complete_samples}
\begin{tabular}{cc}
\hline
$\log(M_{*,\min}/M_\odot)$ & $z_{\max}$ \\
\hline
9.00 & 0.0310 \\
9.25 & 0.0375 \\
9.50 & 0.0460 \\
9.75 & 0.0575 \\
10.00 & 0.0720 \\
\hline
\end{tabular}
\end{table}

\subsection{Complete Samples}
\label{subsec:complete_sample}

Rather than working with a single flux-limited sample, we construct five stellar-mass-complete samples. This serves two purposes. First, if one-halo conformity depends on stellar mass, comparing measurements across complete samples with different stellar-mass limits allows us to test for such variation explicitly. Second, this strategy makes fuller use of the SDSS data by combining lower-redshift complete samples at low mass with higher-redshift complete samples at higher mass, thereby increasing the effective sample size while maintaining well-defined completeness limits.

For a chosen lower stellar-mass threshold $\log(M_{*,\min}/M_\odot)$, we determine the corresponding maximum redshift $z_{\max}$ empirically. Specifically, we consider all galaxies with stellar masses in the interval
\begin{equation}
    \log(M_*/M_\odot) \in
    \left[\log(M_{*,\min}/M_\odot),\,
    \log(M_{*,\min}/M_\odot)+0.1\right].
\end{equation}
For galaxies in this narrow mass slice and within $0.01<z<z_{\max}$, we compute $V_{\max}$ using \texttt{kcorrect v4\_1\_4} \citep{2007AJ....133..734B}. We define $V_{\max}$ as the ratio of the maximum comoving volume over which a galaxy would remain within the SDSS flux limit to the total comoving volume over $0.01<z<z_{\max}$. We then choose $z_{\max}$ such that 95\% of galaxies in this slice satisfy $1/V_{\max}<1.5$, which provides a practical balance between maintaining high completeness and retaining a sufficiently large sample. \added{We have also verified that adopting the more conservative stellar-mass-dependent $z_{\max}$ limits of \citet{2008MNRAS.387...79V} does not change our qualitative conclusions.}

\begin{figure}[t]
    \centering
    \includegraphics[width=0.45\textwidth]{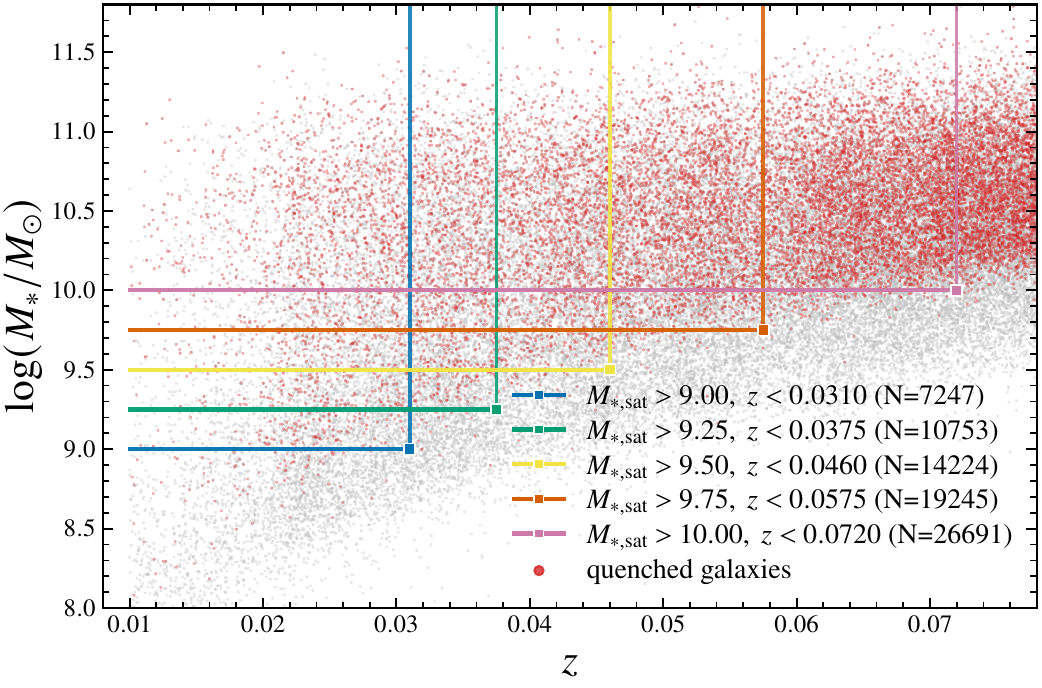}
    \caption{Definition of the five complete samples used in this work. The colored boundaries show the adopted combinations of stellar mass threshold and maximum redshift. Gray points show the parent galaxy sample, and red points denote quenched galaxies, which are more susceptible to incompleteness near the survey limit. $N$ is the total number of satellite galaxies in each sample. The adopted cuts therefore define a practical completeness boundary for the subsequent analysis.}
    \label{fig:sample}
\end{figure}

The resulting five complete samples are listed in Table~\ref{tab:complete_samples} and illustrated in Figure~\ref{fig:sample}. Quenched galaxies are shown explicitly in Figure~\ref{fig:sample} because they are more susceptible to incompleteness near the survey limit and therefore trace the more restrictive completeness boundary. The adopted cuts lie close to the lower envelope of the quenched population, indicating that the resulting completeness limits are neither unnecessarily high nor too close to the obviously incomplete regime. 

For the conformity analysis, we use all satellites in each complete sample with
\begin{equation}
    \log(M_{*,{\rm sat}}/M_\odot) >
    \log(M_{*,\min}/M_\odot),
\end{equation}
and require their central galaxies to satisfy
\begin{equation}
    \log(M_{*,{\rm cen}}/M_\odot) >
    \max\!\left[9.5,\,\log(M_{*,\min}/M_\odot)\right].
\end{equation}
The second condition is imposed because the ML halo masses are only available for groups with central stellar masses above $10^{9.5} M_\odot$. Since satellites are by construction less massive than their centrals, this defines the final analysis sample used throughout the paper.

All galaxies are assigned equal weight in the main analysis.

\subsection{Conformity Statistics}
\label{subsec:statistics}

We first measure one-halo conformity in the satellite-level form commonly adopted in the previous literature \citep[e.g.][]{2006MNRAS.366....2W,2013MNRAS.430.1447K}. Halo masses are binned into six uniform bins over the range
\begin{equation}
    11.5 \le \log(M_{\rm halo}/M_{\odot}) < 14.5
\end{equation}
with a bin width of 0.5 dex.\footnote{For the two AM halo-mass estimates, the Yang et al.\ group catalog has an effective low-mass truncation around $\log(M_{\rm halo}/M_{\odot}) \simeq 11.8$. As a result, the lowest AM bin is not populated all the way down to 11.5, although we keep the same nominal binning for consistency across the three halo-mass estimates.}

In each halo-mass bin, we compute the median sSFR of satellites around star-forming centrals and that of satellites around quenched centrals, and define the satellite-level conformity signal as
\begin{equation}
\begin{split}
\Delta(\log ({\rm sSFR}))_{\rm sat}
    =\,& {\rm median}(\log ({\rm sSFR}))_{{\rm sat}|{\rm SF\,cen}} \\
      &- {\rm median}(\log ({\rm sSFR}))_{{\rm sat}|{\rm Q\,cen}} .
\end{split}
\end{equation}

Because each satellite enters this statistic with equal weight, groups with more members naturally receive greater statistical weight in the satellite-level measurement. Since one-halo conformity is physically a halo-scale phenomenon, we also construct a halo-level statistic, in which each group contributes more equally to the final signal. For this purpose, we first reduce each group to a single summary value by taking the median sSFR of all satellites belonging to that halo, and then compare this halo-level satellite summary between star-forming-central and quenched-central halos in the same halo-mass bin:
\begin{equation}
\begin{split}
\Delta(\log ({\rm sSFR}))_{\rm halo}
    =\,& {\rm median}\!\bigl(\log ({\rm sSFR})_{\rm halo}\bigr)_{{\rm SF\,cen}} \\
      &- {\rm median}\!\bigl(\log ({\rm sSFR})_{\rm halo}\bigr)_{{\rm Q\,cen}},
\end{split}
\end{equation}
where
\begin{equation}
    \log ({\rm sSFR})_{\rm halo}
    =
    {\rm median}(\log ({\rm sSFR}))_{\rm sat\ in\ halo}.
\end{equation}

We carry out both measurements for all three halo-mass estimates: ML, mass-ranking AM, and luminosity-ranking AM.

\added{We use sSFR rather than a binary quenched fraction as our primary measure of satellite star-formation activity. The continuous sSFR retains information that is lost in a binary classification and can be applied consistently to both our satellite-level and halo-level analyses; in contrast, the quenched fraction of an individual halo becomes highly discrete for low-richness systems. Moreover, the satellite sSFR distributions over the parameter ranges considered here do not exhibit a pervasive, strongly separated bimodality that would make a binary statistic clearly preferable. We therefore adopt sSFR as our conformity statistic. }

\subsection{Uncertainties and Constant-Amplitude Fits}
\label{subsec:uncertainty}

Uncertainties are estimated with bootstrap resampling, using halo/group as the resampling unit in all cases. This avoids treating multiple satellites within the same halo as statistically independent. We use 1000 bootstrap realizations and take the 16th and 84th percentiles of the bootstrap distribution as the approximate $1\sigma$ uncertainty range.

To summarize each $\Delta(\log {\rm sSFR})$--$M_{\rm halo}$ relation with a single number, we fit a constant across the halo-mass bins. For measurements $(y_i,\sigma_i)$ in different bins, the best-fitting constant, denoted by $\langle \Delta(\log {\rm sSFR}) \rangle$, is
\begin{equation}
    \langle \Delta(\log {\rm sSFR}) \rangle
    =
    \frac{\sum_i y_i/\sigma_i^2}{\sum_i 1/\sigma_i^2},
\end{equation}
with uncertainty
\begin{equation}
    \sigma_{\langle \Delta \rangle}
    =
    \left(\sum_i \frac{1}{\sigma_i^2}\right)^{-1/2}.
\end{equation}

\section{Results}
\label{sec:results}

\begin{figure*}[t]
    \centering
    \includegraphics[width=\textwidth]{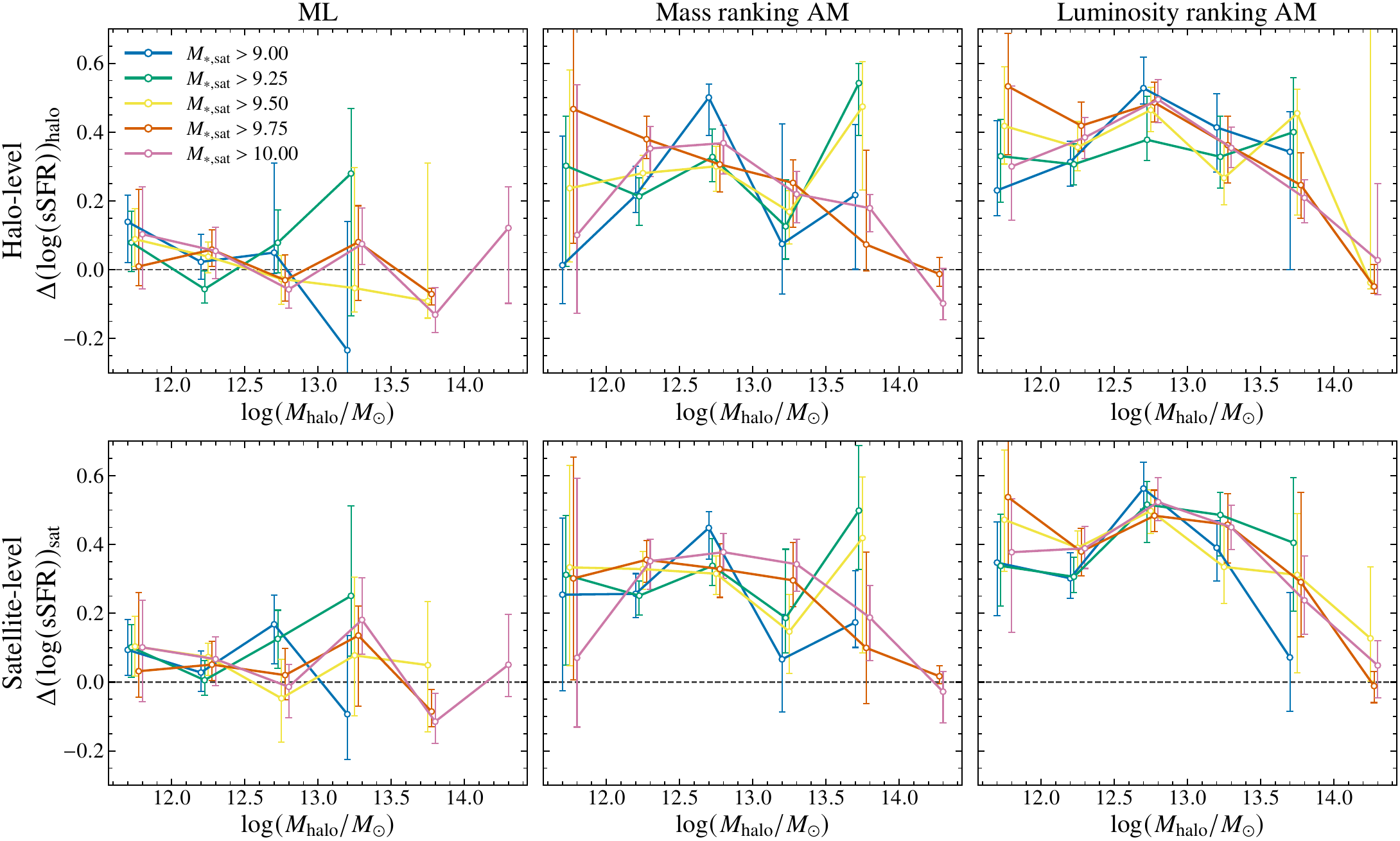}
    \caption{Raw one-halo conformity signals measured from the five complete samples. The upper row shows the halo-level statistic, while the lower row shows the satellite-level statistic. From left to right, the three columns use halo masses derived from the ML method, the mass-ranking AM method and the luminosity-ranking AM method, respectively. Different colors correspond to the five complete samples defined in Figure~\ref{fig:sample}. Error bars show the bootstrap uncertainties, and the dashed horizontal lines mark zero signal.
}
    \label{fig:original_signal}
\end{figure*}

\begin{figure*}[]
    \centering
    \includegraphics[width=\textwidth]{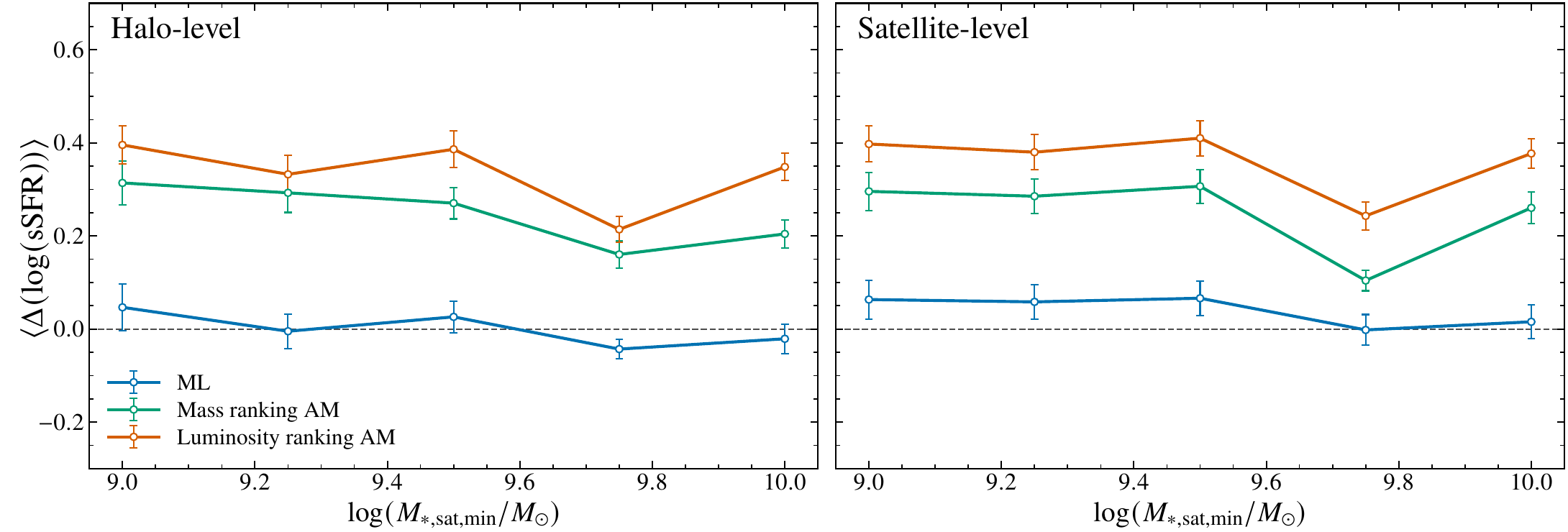}
    \caption{Best-fitting constant amplitudes obtained by fitting a horizontal line to each raw conformity curve shown in Figure~\ref{fig:original_signal}. The left panel shows the halo-level results and the right panel shows the satellite-level results. Blue, green, and orange lines correspond to the ML, mass-ranking AM, and luminosity-ranking AM halo masses, respectively. Error bars indicate the $1\sigma$ uncertainties of the fitted constant amplitudes. The fitted amplitudes depend systematically on the adopted halo-mass method.
}
    \label{fig:fitted_signal}
\end{figure*}

We begin with the raw one-halo conformity signals shown in Figure~\ref{fig:original_signal}. In each panel, the five colored curves correspond to the five complete samples defined in Section~\ref{subsec:complete_sample}, while the three columns compare the three halo-mass estimates: ML, mass-ranking AM, and luminosity-ranking AM. The upper row shows the halo-level statistic, and the lower row shows the satellite-level statistic.

A clear result from Figure~\ref{fig:original_signal} is that the inferred conformity signal depends strongly on the adopted halo-mass estimate. For both the halo-level and satellite-level statistics, the two AM methods generally yield positive conformity signals over a broad range of halo mass, with the luminosity-ranking AM method giving the strongest signal and the mass-ranking AM method giving a somewhat weaker but still clearly positive signal. By contrast, the ML-based curves are much closer to zero. 


The raw conformity curves in Figure~\ref{fig:original_signal} show no
compelling evidence for a strong halo-mass dependence
\added{over most of the halo-mass range} within the measurement
uncertainties. \added{The main exception occurs at the high-mass end, where
the AM-based conformity signal becomes weaker and the measurements are
consistent with no conformity within the uncertainties and broadly
consistent with the ML-based results. This convergence at high halo masses
is reassuring, since abundance-matching halo masses are expected to be more
reliable for richer and more massive groups.} We therefore summarize each curve by fitting a constant across the halo-mass bins. This horizontal-line fit provides a compact description of the overall conformity amplitude and facilitates a cleaner comparison among the five complete samples. The resulting best-fitting amplitudes are shown in Figure~\ref{fig:fitted_signal}. The same qualitative pattern remains when each curve is summarized by a horizontal-line fit: both AM methods yield systematically positive amplitudes, whereas the ML-based amplitudes remain close to zero.

Figure~\ref{fig:fitted_signal} further shows that the fitted amplitudes vary only weakly across the five complete samples, indicating that the inferred signal does not depend strongly on the adopted satellite stellar-mass threshold. We therefore summarize each combination of signal definition and halo-mass estimate by taking the mean fitted amplitude across the five complete samples, together with the mean of the corresponding uncertainties as a representative error.

For the halo-level sSFR signal, the representative amplitudes are $+0.3790 \pm 0.0395$, $+0.2333 \pm 0.0396$, and $+0.0008 \pm 0.0347$ for the luminosity-ranking AM, mass-ranking AM, and ML halo masses, respectively. These correspond to approximate significances relative to zero of $9.6\sigma$, $5.9\sigma$, and $0.02\sigma$. For the satellite-level sSFR signal, the corresponding representative amplitudes are $+0.3989 \pm 0.0370$, $+0.2687 \pm 0.0378$, and $+0.0403 \pm 0.0371$, corresponding to $10.8\sigma$, $7.1\sigma$, and $1.1\sigma$, respectively.

Taken together, Figures~\ref{fig:original_signal} and \ref{fig:fitted_signal} show that the strong one-halo conformity signal inferred from the AM halo masses is not reproduced when the same analysis is repeated using the ML halo masses. In our analysis, the ML-based signal is weak in both statistics, and is especially weak at the halo level, where it is fully consistent with zero. This suggests that the strong conformity signal obtained with AM halo masses is likely driven largely by errors in the AM halo-mass estimates.

\section{Robustness to Group-finder Systematics}
\label{sec:groupfinder}

\added{

\begin{figure*}[t]
    \centering
    \includegraphics[width=\textwidth]{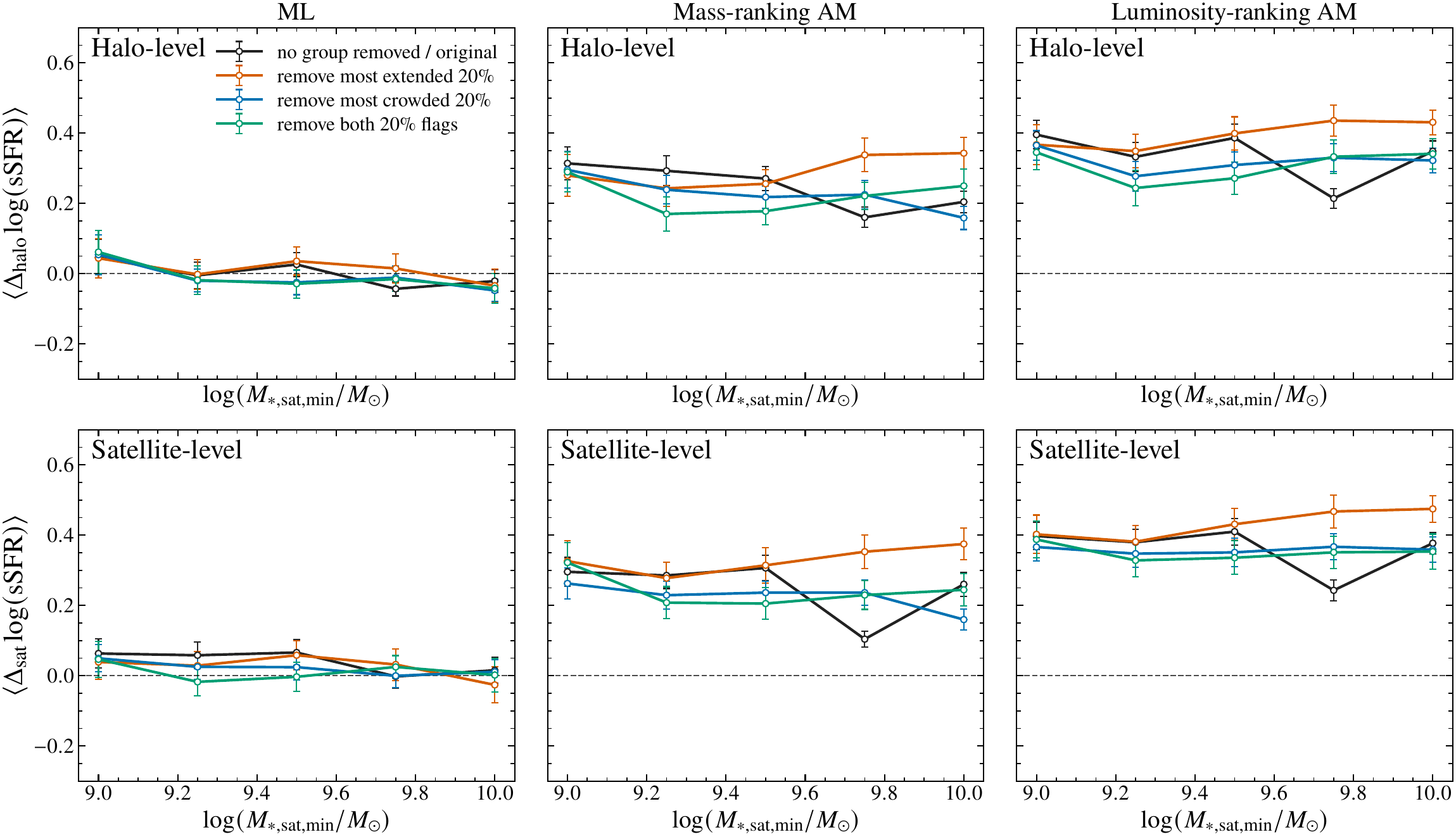}
    \caption{
    Robustness of the one-halo conformity measurements to group
    configurations that are potentially more susceptible to group-finder
    systematics. The top and bottom rows show the halo-level and
    satellite-level measurements, respectively. The left, middle, and right
    columns use the ML, stellar-mass-ranking abundance-matching, and
    luminosity-ranking abundance-matching halo-mass estimates. Black points
    show the original measurements. Orange and blue points show the results
    after removing the most extended 20\% and the most crowded 20\% of groups,
    respectively, while green points show the results after both populations
    are removed. }
    \label{fig:group_finder}
\end{figure*}

In addition to uncertainties in halo-mass estimation, the measurement of
one-halo conformity may also be affected by systematics associated with the
group-finding procedure itself. In particular,
\citet{2015MNRAS.452..444C} showed using mock catalogs that incorrect
group-membership assignments can affect inferred conformity statistics.
Two important failure modes are ``fusing'', in which galaxies associated
with distinct underlying halos are assigned to the same catalog group,
and ``fracturing'', in which galaxies belonging to the same underlying halo
are assigned to multiple catalog groups. Since improving the halo-mass
estimator cannot correct an incorrect group-membership assignment, we perform
an additional observational test of the sensitivity of our results to these
group-finder systematics.

Although individual fused or fractured systems cannot be identified
unambiguously in the observational catalog without knowledge of the true
halo membership, we identify group configurations that are expected to be
more susceptible to these effects. First, we define the spatial extent of a
group as
\begin{equation}
R_{\rm ext}\equiv\frac{R_{p,\max}}{R_{\rm vir}},
\end{equation}
where $R_{p,\max}$ is the maximum projected distance of its satellites from
the catalog group center and
$R_{\rm vir}$ is the virial radius derived from the ML
halo-mass estimate. We classify the 20\% of groups with the largest
$R_{\rm ext}$ as extended groups. Such extended satellite
distributions are expected to be more susceptible to potential fusing by
the group finder.

Second, for each group $i$, we identify the nearest neighboring group $j$
in projected separation among those satisfying
$|\Delta v_{\rm los}|<1000~{\rm km\,s^{-1}}$, and define
\begin{equation}
D_{\rm NN}\equiv
\frac{R_{p,ij}}
{\max(R_{{\rm vir},i},R_{{\rm vir},j})},
\end{equation}
where $R_{p,ij}$ is the projected separation between groups $i$ and $j$,
and $R_{{\rm vir},i}$ and $R_{{\rm vir},j}$ are their respective virial
radii. We classify the 20\% of groups with the smallest $D_{\rm NN}$ as
crowded groups. Such closely neighboring systems are expected to be more
susceptible to potential fracturing by the group finder. We have also tested
alternative selection thresholds and find that our conclusions are unchanged.

We repeat the conformity analysis after removing the extended groups, the
crowded groups, and both populations simultaneously.
Figure~\ref{fig:group_finder} shows the resulting measurements. For the ML
halo masses, the conformity signal remains close to zero at both the halo
and satellite levels under all of these selections. In contrast, the
positive conformity signals obtained with both abundance-matching estimators
persist. Thus, removing group configurations that are potentially more
susceptible to group-finder errors does not materially alter the contrast
between the ML and abundance-matching measurements.

These observational selections cannot identify or eliminate all cases of
fracturing and fusing, and therefore do not rule out group-finder systematics
completely. Nevertheless, the stability of the measurements indicates that
membership ambiguities associated with the most extended or crowded systems
are unlikely to be large enough to alter our main conclusion. Moreover, in
several of the cleaned samples, the already weak ML-based conformity signal
moves closer to zero. This behavior is qualitatively consistent with the
result of \citet{2015MNRAS.452..444C} that group-finding errors may contribute
a weak positive conformity signal, rather than accounting for the
substantially stronger signal obtained with the abundance-matching halo
masses.
}

\section{Summary}

One-halo galactic conformity is the tendency for satellites in halos with quenched centrals to have lower star-formation activity than those in halos with star-forming centrals at fixed halo mass. It is an important probe of the galaxy--halo connection and halo-wide quenching processes, such as halo quenching and AGN feedback. However, its detection remains controversial, because conformity must be measured at fixed halo mass and therefore requires reliable halo-mass estimates.

In this Letter, we addressed this issue using SDSS DR7 with five stellar-mass-complete samples and three halo-mass estimates: a new machine-learning (ML) estimate and two conventional abundance-matching (AM) estimates based on mass ranking and luminosity ranking. We measured the signal with both a satellite-level statistic, which compares the median sSFR of satellites around quenched and star-forming centrals within the same halo-mass bins, and a halo-level statistic, which first assigns each halo the median sSFR of its satellites and then compares quenched- and star-forming-central halos at fixed halo mass.

Our results show that the two AM halo-mass estimates yield strong positive conformity signals, broadly consistent with previous AM-based measurements in the literature. These signals, however, are not reproduced when the same analysis is performed with the ML halo masses. For the halo-level statistic, the representative AM-based signals are $+0.38 \pm 0.04$ dex for the luminosity-ranking AM halo masses and $+0.23 \pm 0.04$ dex for the mass-ranking AM halo masses, corresponding to detections relative to no conformity at about $10\sigma$ and $6\sigma$, respectively. In contrast, the ML result is consistent with no conformity, $+0.00 \pm 0.03$ dex. For the satellite-level statistic, the corresponding signals are $+0.40 \pm 0.04$ dex, $+0.27 \pm 0.04$ dex, and $+0.04 \pm 0.04$ dex for the luminosity-ranking AM, mass-ranking AM, and ML halo masses, respectively.


The measured signal varies only weakly across the five complete samples, indicating little dependence on the adopted satellite stellar-mass threshold. \added{We further find that the contrast between the ML- and AM-based measurements is robust to removing group configurations that are potentially more susceptible to group-finder systematics: the ML-based signal remains close to zero, while the positive AM-based signals persist.}
\added{Taken together, with halo-mass estimates that reproduce the current weak-lensing constraints on the stellar mass--halo mass relations of star-forming and quenched central galaxies, we find no detectable one-halo conformity signal in the present SDSS sample.} The strong signal inferred from AM halo masses is therefore likely driven largely by errors in the AM halo-mass estimates. The absence of a detectable ML-based signal provides an important observational constraint on the galaxy--halo connection and halo-wide quenching models. Combined with the absence of an intrinsic two-halo conformity signal found by \citet{2023MNRAS.523.1268W}, our result suggests that there is no clear evidence for intrinsic galactic conformity in the present low-redshift observations.

\begin{acknowledgments}
We thank the anonymous referee for their careful reading and constructive comments, which helped improve the manuscript. Y.P. and D.Z. acknowledge support from the National Natural Science Foundation of China (NSFC) under grant Nos. 12125301, 12192220, and 12192222. Y.P. also acknowledges support from the New Cornerstone Science Foundation through the XPLORER PRIZE. K.W. acknowledges support from the Science and Technologies Facilities Council (STFC) through grant ST/X001075/1. Y.J.W. acknowledges supports by National Natural Science Foundation of China (Project No. 12403019) and Jiangsu Natural Science Foundation (Project No. BK20241188). J.D. acknowledges the support of National Science Foundation of China (NSFC) grant Nos. 12303010. This work is extensively supported by the High-performance Computing Platform of Peking University, China.
\end{acknowledgments}

\bibliography{sample701}{}
\bibliographystyle{aasjournalv7}



\end{document}